\documentclass[3p, twocolumn]{elsarticle}

\usepackage{array}
\usepackage{subfigure}
\usepackage{caption}
\usepackage{amssymb}
\usepackage{graphicx}
\usepackage{amsmath}

\newcolumntype{x}[1]{>{\centering\arraybackslash\hspace{0pt}}p{#1}}

\usepackage[group-separator={,}]{siunitx}

\graphicspath{ {./Figures/} }
\usepackage{hyperref}

\journal{Nuclear Instruments and Methods in Physics Research Section A}

\begin{document}
\begin{frontmatter}

\title{Optical Calibration System for the LUX-ZEPLIN (LZ) Outer Detector}

\author[1]{W.~Turner\corref{cor1}}\ead{w.turner@liverpool.ac.uk}
\author[1]{A.~Baxter}
\author[1,2]{H.~J.~Birch}
\author[1,3]{B.~Boxer}
\author[1]{S.~Burdin}
\author[1]{E.~Fraser}
\author[1]{A.~Greenall}
\author[1]{S.~Powell}
\author[1]{P.~Sutcliffe}

\cortext[cor1]{Corresponding author}

\address[1]{University of Liverpool, Oliver Lodge Laboratory, Oxford Street, Liverpool, L69 7ZE, UK}
\address[2]{Department of Physics, University of Michigan, Ann Arbor, Michigan 48109, USA}
\address[3]{STFC  Rutherford  Appleton  Laboratory  (RAL),  Didcot,  OX11  0QX,  UK}

\begin{abstract}
The LUX-ZEPLIN experiment will search for dark matter particle interactions with a detector containing a total of 10 tonnes of liquid xenon. Surrounding the liquid xenon cryostat is an outer detector veto system with the primary aim of vetoing neutron single-scatter events in the liquid xenon that could mimic a weakly interacting massive particle (WIMP) dark matter signal. The outer detector consists of approximately 17~tonnes of gadolinium-loaded liquid scintillator confined to 10 acrylic tanks surrounding the cryostat and $\num{228000}$ litres of water as the outermost layer. It will be monitored by 120 inward-facing 8-inch photomultiplier tubes. An optical calibration system has been designed and built to calibrate and monitor these photomultiplier tubes allowing the veto system to reach its required efficiency and thus ensuring that LUX-ZEPLIN meets its target sensitivity. 
\end{abstract}


\end{frontmatter}



\section{Introduction}
LUX-ZEPLIN (LZ) is a dark matter direct detection experiment with liquid xenon (LXe) as the target material. Its projected sensitivity (90\% C.L.) for WIMP-nucleon spin-independent interactions reaches $1.4 \times 10^{- 48}$~cm$^2$ for $40$~GeV/c$^{2}$ WIMPs \cite{lzSens} \cite{LZNIMA}. The core of LZ is a dual-phase xenon time projection chamber (TPC) with an active mass of 7 tonnes which will be the largest detector of its kind. The TPC is surrounded by an outer detector (OD) which is fundamental in reaching the desired sensitivity. In the TPC, signals produced by neutrons are indistinguishable from those of WIMPs on an event-by-event basis; the purpose of the OD is to veto these neutrons~\cite{LZTDR}. LZ is in the final stages of commissioning ${\sim}$1.5~km (${\sim}$4.5~km water equivalent) underground at the Sanford Underground Research Facility (SURF).

\begin{figure*}[t]
  \centering
  \includegraphics[width=0.75\linewidth]{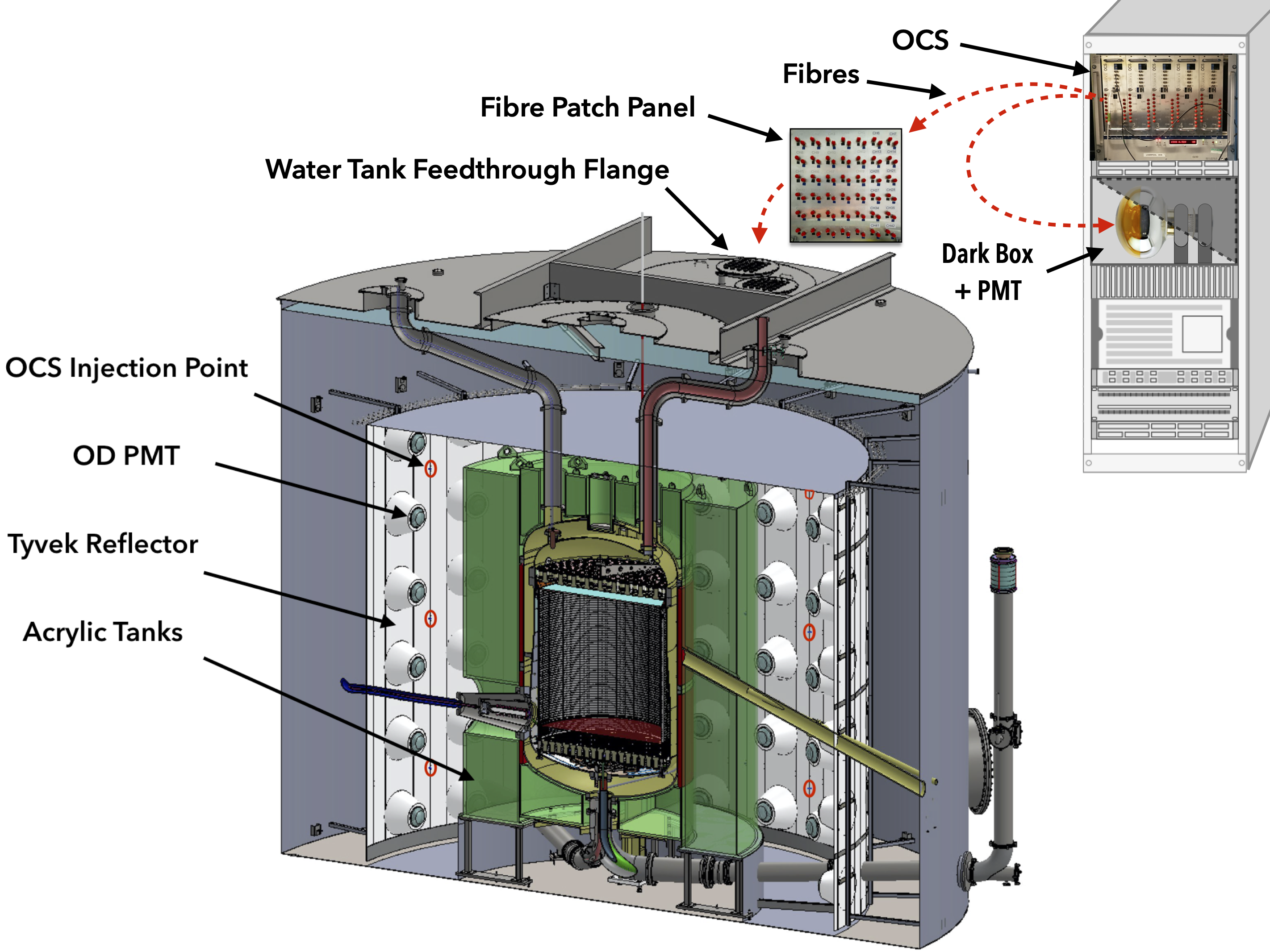}
  \caption{Cutaway drawing of the LZ detector systems. The LXe TPC in the centre is surrounded by the scintillator tanks (green) and light collection system (white), all housed in a large water tank (blue-grey). The positions of some of the injection points for the OD OCS are highlighted by red circles. The route the produced light takes to reach the water tank and the monitoring PMT housed in the dark box is shown as red dashed line. Schematic of the LZ detector is taken from LZ TDR \cite{LZTDR}.}
  \label{fig:ocsTotalDiagram}
\end{figure*} 

The OD and a LXe `skin' detector compose the LZ veto system which together form a active shield for radiation entering or escaping from the TPC, as shown in Fig.~\ref{fig:ocsTotalDiagram}. Schematic of the LZ detector taken from LZ TDR \cite{LZTDR}. The LXe `skin' detector is located between the TPC and the inner cryostat wall and is principally used to veto MeV-scale gamma rays. The OD forms a near hermetic seal around the cryostat and is composed of a  segmented acrylic tank housing $17$ tonnes of linear alkylbenzene based gadolinium-loaded liquid scintillator and is surrounded by $\num{228000}$ litres of deionised water. The segmented acrylic tank and housed scintillator are refereed to collectively as the scintillator tanks. The primary function of the scintillator tanks is to veto neutrons that have interacted in the TPC, predominantly via neutron capture on gadolinium \cite{GdLSDayaBay}. Light generated in the OD is subsequently collected by $120$ inward facing 8-inch Hamamatsu R5912 photomultiplier  tubes (PMTs) \cite{ODPMT}. To maximise light collection efficiency, there is a Tyvek\textsuperscript{\textregistered} curtain behind, above and below the PMTs, and a layer of Tyvek\textsuperscript{\textregistered} surrounding the cryostat \cite{LZTDR}.

The main requirement to the OD is to provide a veto efficiency to the neutrons scattered in the TPC greater than 95\% with less than 5\% dead time at the effective threshold of 200~keV (to be above the end of the $^{14}$C beta spectrum.) \cite{lzSens}. This requires understanding the OD optical model and PMT stability at the $\sim$1\% level. An Optical Calibration System (OCS) has been designed and built to validate and monitor OD optical properties and calibrate the OD PMTs at the required level. Though the expected OD threshold is 200~keV, the OCS was designed to calibrate the OD down to 150~keV.

Section~\ref{OCS_Req} of this article describes the design choices made to ensure the OCS meets the given requirements, while Sec.~\ref{OCS_sysOverview} outlines the hardware developed to meet these requirements as well as the hardware selection criteria. Section~\ref{OCS_calibration} presents the calibration procedure and Sec.~\ref{OCS_performance} shows the final system performance. Additionally, Sec.~\ref{OCS_RadiationMeasurments} presents the radioassaying results of the OCS components and how the observed rates compare to that of the outer detector.


\section{System Requirements} \label{OCS_Req}
The OCS must be capable of executing a number of routines to monitor the OD. The requirements underpinning these routines informed the design of the OCS. These requirements are summarised in Tab.~\ref{tab:requirements} and described below.

Verification of the OD PMTs response is prerequisite to all other monitoring techniques. Full geometry simulations were carried out using a GEANT4 based simulation framework developed by the LZ collaboration, BACCARAT~\cite{LUXSim}~\cite{LZ_SIMS} \cite{AGOSTINELLI2003250}. This demonstrated that the response of OD PMTs can be monitored via the repeated injection of $\num{1000}$~photons from the central row of fibres. These injections yield a high proportion of 1-photon pulses collected by each of the 120 PMTs. The BACCARAT simulations show that the average pulse area resulting from 1 photoelectron escaping the photocathode of a PMT can be determined with an uncertainty of $\sim$1\%, given enough statistics. Therefore, any change in PMT response above 1\% would be detected.

\begin{table*}[t]
    \centering
    \small
    \caption{Requirements for the OD OCS with the reason for each requirement. Based on LZ TDR requirements for the OD \cite{LZTDR}.}
    \begin{tabular} {|m{0.4\textwidth}|m{0.55\textwidth}|}
        \hline
        Requirement & Reason \\
        \hline
          
        Ability to inject $\num{700}$\,--\,$\num{50000}$~photons per channel. & To be able to test the threshold energy of 150~keV ($\num{1350}$~photons) as well as being able to match the signals of the calibration sources which can go up to about 2.6~MeV ($\num{20000}$~photons).\\ \hline
        
        Total number of photons: 1M~photons globally. & To mimic the average muon signal. \\ \hline
        
        Variation between injected pulses: $\lesssim 10\% $. & Should be no worse than the statistical variation of the OD. \\ \hline
        
        Variation between calibrations: $\lesssim 100$~photons at 150~keV. & Need to be able to keep the threshold of the OD stable at around 150~keV. \\ \hline
        
        Light intensity precision: $\num{100}$~photon precision for $\num{700}$--$\num{2000}$ injected photons; $\num{1000}$ photon precision for $\num{5000}$--$\num{10000}$ injected photons. & To allow scanning over the OD threshold region and simplify the match to expected background and calibration sources.\\ \hline
        
        Pulse width: $< 20$~ns. & Should be shorter than the amplifier shaping time of the DAQ electronics (30~ns).\\ \hline
        
        Injected light wavelength: $430$\,--\,$450$~nm, $450$\,--\,$460$~nm, $365$\,--\,$390$~nm. & Matches the peak wavelength and quantum efficiency of the OD PMTs and the scintillation light from the GdLS.\\ \hline
        
        Pulsing frequency: up to $10$~kHz. & Ensure calibrations use the full readout bandwidth.\\ \hline
        
        Alignment: $<5^{\circ}$ misalignment. & To limit variation in the number of photons in each PMT. This value is an acceptable upper limit to satisfy this requirements.\\ \hline
        
        Radioactivity: $< 5\%$ of the total OD rate. & To ensure the background rates in the detector are kept to a minimum.\\ \hline
    \end{tabular}
    \label{tab:requirements}
\end{table*}

To sufficiently monitor the OD PMT response at the 150~keV threshold (which corresponds to about 20 detected photons and $\sim$1300~photons) a scan must be made over this point, therefore the minimum number of photons injected by a single fibre needs to be about 700. Around 1 million photons are required to study afterpulsing from energetic cosmic muons; this is the maximum requirement for the OCS. To achieve this, each channel must be capable of injecting at least $\num{50000}$~photons which when pulsed in synchrony with each other will reach 1 million photons.

When monitoring the light collection efficiency at the energy scale of the calibration source, variation in the number of photons collected from an injection needs to be less than the variation in the calibration source to avoid widening subsequent calibration peaks. Each pulse produced from the OCS is monitored by the rack mounted PMT. This allows for pulse-to-pulse calibration. BACCARAT simulations show the resolution value of the peaks resulting from the calibration sources to be ${\sim}$12\%. A target of less than 10\% variation in the number of photons detected after an injection was hence set. 

To have sufficient resolution to monitor the threshold region and the energies deposited by the calibration sources, as given in Tab.~\ref{tab:calibrationSources}, the OCS injections must have a precision of 100 photons for injections of $\num{700}$--$\num{2000}$~photons and a precision of $\num{1000}$~photons between $\num{5000}$--$\num{10000}$~photons. High intensity signals are not connected to the calibration sources and therefore the precision is less of a concern. 

The requirement outlining the wavelengths of light to inject is not only to ensure the injected wavelength matches the peak wavelength and quantum efficiency of the OD PMTs but to also ensure scintillator degradation can be detected. The absorption length in the scintillator decreases significantly for wavelengths below 420~nm, this region shifts to higher wavelengths as the scintillator degrades \cite{ScintillatorPerform}. Injections of 435~nm and 450~nm photons from the same injection point directly below the scintillator tanks will allow monitoring of this degradation. A similar relationship between the absorption of UV light and the degradation of the acrylic housing of the scintillator exists and will be monitored via the same method using one injection point under the acrylic tank lip with fibres connected to $\num{390}$~nm and $\num{435}$~nm LEDs. 

The last three requirements in Tab.~\ref{tab:requirements} are more practical in nature. To make use of the full 4~kHz bandwidth of the DAQ~\cite{LZTDR}, the OCS should be able to complete injection at a rate faster than this. The maximum rate depends on occupancy and it will be determined experimentally. A target requirement of 10~kHz was chosen for the OCS electronics as this would allow collection of suitable statistics in a short amount of time. To ensure that realistic comparisons can be made with the simulations, the misalignment of fibres in the OD needs to be less than 5$^{\circ}$.

\begin{table*}[t]
    \centering
    \small
    \caption{Sources used to calibrate the OD with the energy deposited in the OD and the average number of photons produced and detected listed. To go from the energy deposited to photons produced a constant of $\num{9000}$~photons/MeV have been used, and to calculate the approximate number of photons detected the average light collection efficiency around $7\%$ and mean PMT quantum efficiency around $25\%$ were taken into account \cite{LZ_SIMS} \cite{LZTDR}.}
    \begin{tabular} {|m{0.15\textwidth}|m{0.20\textwidth}|m{0.22\textwidth}|m{0.22\textwidth}|}\hline
             Source           & Energy Deposited   & \# Photons Produced & \# Photons Detected \\ \hline 
             $^{22}$Na        & 511 keV         & $\num{4600}$           & $\num{64}$          \\
             $^{22}$Na        & 1275 keV        & $\num{11500}$          & $\num{160}$          \\
             DD               & 2200 keV        & $\num{19800}$          & $\num{277}$          \\
             $^{228}$Th       & 2615 keV        & $\num{23500}$          & $\num{330}$         \\
             Cosmic-ray Muon  & 0.1 - 1 GeV     & $\num{1000000}$+       & $\num{20000}$+         \\ \hline
    \end{tabular}

    \label{tab:calibrationSources}
\end{table*}

Calculation of the neutron veto efficiency is reliant on accurate optical modelling of the OD in BACCARAT simulations. This optical model can be tuned and verified by comparison of light collections in LZ simulation and experimental data using processes which produce predictable light output: cosmic ray muons passing through the OD, calibration sources, and OCS injections. The calibration sources, shown in Tab.~\ref{tab:calibrationSources}, allow absolute calibration of the OD connecting energy depositions in the liquid scintillator with the number of detected photons. These signals will be used as reference points for the OCS. Table~\ref{tab:calibrationSources} uses the energy deposited in the OD from each source to determine the approximate number of photons emitted using $\num{9000}$~photons/MeV, as well as taking into consideration the average light collection efficiency of approximately $7\%$ and mean PMT quantum efficiency about $25\%$ to determine the approximate photons detected \cite{LZ_SIMS}\cite{LZTDR}. The OCS will also be used for injections of light at intermediate and extended intensities, and in regions of the OD which are not reachable by the calibration sources. Given that the OCS calibration is much faster than the calibration with radioactive sources, the OCS calibrations will be performed more frequently.


\section{System Overview} \label{OCS_sysOverview}
The OCS will use duplex optical fibres to inject controlled pulses of light produced by LEDs into the OD at 35 locations. Of these, 30 locations are distributed evenly around the water tank (10 azimuthal positions at 3 heights) allowing for good coverage of the detector. Additionally, four injection points are located beneath the four side scintillator tanks in order to probe the scintillator quality, and one injection point beneath one of the side scintillator tanks to probe the acrylic quality.

\begin{figure*}
    \centering
    \includegraphics[width=0.75\linewidth]{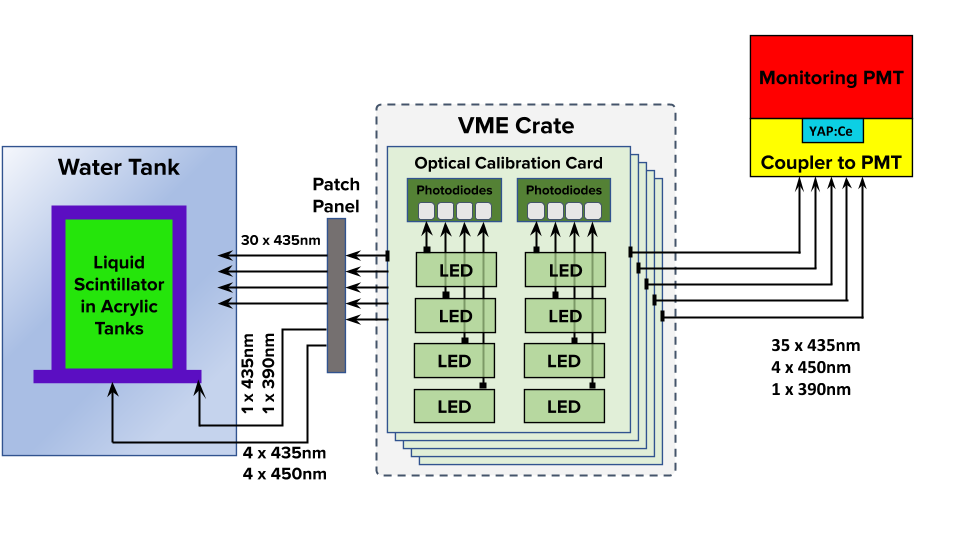}
    \caption{A diagram showing an overview of the Optical Calibration System, with eight LED pulsers on one Optical Calibration Card (OCC) and five Optical Calibration Cards in the VME crate. Lines with arrows show fibre routes with labels representing numbers of fibres from LEDs with corresponding wavelengths.}
    \label{fig:sysOverview}
\end{figure*}

\begin{figure*}
    \centering
    \includegraphics[width=0.99\linewidth]{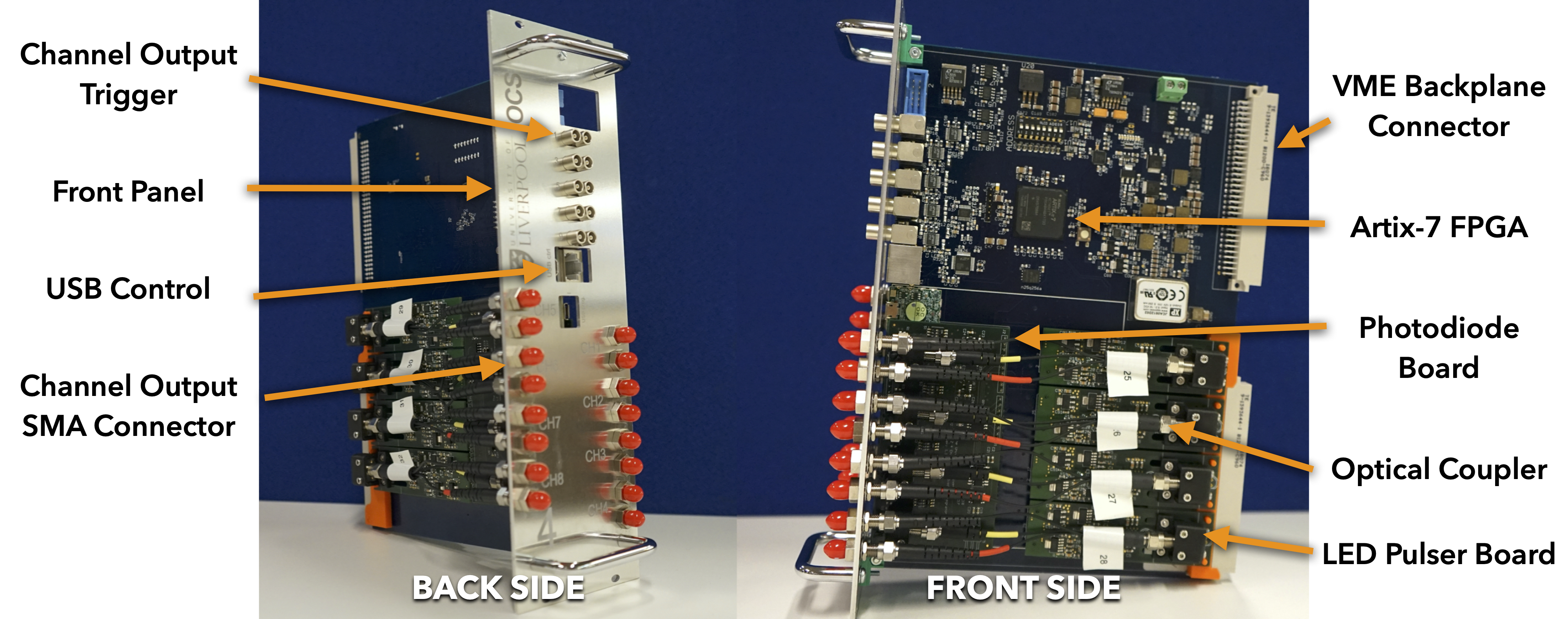}
    \caption{A picture of an Optical Calibration Card with the key components labelled.}
    \label{fig:OCC}
\end{figure*}

The OCS electronics system consists of five Optical Calibration Cards (OCC). Each card consists of a custom-made Field Programmable Gate Array (FPGA) motherboard, which houses eight LED pulser boards as well as two photo-diode boards. Light from each LED is fed to two outputs on the front panel and to a photo-diode input via a custom made 3-way optical coupler. The components of an OCC can be seen in Figs.~\ref{fig:sysOverview} and \ref{fig:OCC}.

Light is fed from the rack housing the OCS electronics to the patch panel via 40 optical fibres 15~m in length. From the patch panel, they are coupled via SMA-SMA connectors to 21~m long duplex fibres that lead to the injection points inside the water tank. To enter the water tank, the fibres pass through an air-tight and light-tight feedthrough flange located on top of the water tank. For the injection points located around the sides of the detector, only one of the cores of the duplex fibre will be used to inject light into the detector. The other core will be available for potential future upgrades or in case of damage to the first core. The upward facing fibres at the bottom of the scintillator tanks will inject light of two different wavelengths via the two different cores, as mentioned in Sec.~\ref{OCS_Req} and described in detail in Sec.~\ref{SYS_Electronics}.

Precise monitoring of the light intensity is carried out in two ways; firstly, via the FPGA board controlled photo-diode boards and, secondly, via an 8~inch Hamamatsu R5912 PMT installed in a rack-mounted dark box close to the OCS electronics. This PMT is identical to those used in the OD. The stability of this PMT is also monitored, using a YAP:Ce pulser unit made by Scionix \cite{YapCe} which produces light pulses corresponding to 5k photoelectrons with a rate of 20~counts/s.

The OCS will receive the pulse configuration from the LZ Slow Control system which will also store the required OCS calibration database. Between the Slow Control system and the OCS crate is the OCS controller. The controller will have two purposes: to translate the Slow Control system commands into the required FPGA board register values and read back the board temperatures and photo-diode values as well as the FPGA status messages to display in the Slow Control.

\subsection{Electronics} \label{SYS_Electronics}
Each FPGA motherboard sends Low Voltage Digital Signals (LVDS) to the LED pulser boards which are then converted into a single ended Transistor-Transistor Logic (TTL) signal. This signal then drives the base of a common collector transistor circuit which changes the voltage on the cathode of the LED. When the LED is turned off, the anode is held at $-$1~V. Reverse bias ensures that the LED does not produce an unintentional light emission which could lead to fake signals in the OD. When the LED is fired, the transistor is turned off and the cathode voltage goes to a nominal low voltage forward biasing the LED. The number of photons emitted per pulse is controlled by the time width of the LVDS.

High output violet blue LEDs, BRITE-LED Optoelectronics BL-LBVB5N15C, of wavelength 435~nm were selected to match the scintillation light from the GdLS \cite{GdLSDayaBay} and the peak quantum efficiency of the OD PMTs. Quality control checks were completed to ensure that the LEDs selected had similar intensities in order to minimise channel to channel variation. Of the 127 LEDs tested, 35 were selected for use (30 for side injections and 5 for injections beneath the scintillator tanks). To select the LEDs, first a subset of 60 LEDs of similar capacitance were chosen. Then, the intensities of the subset LEDs as measured by the test stand PMT (Hamamatsu H10721-20) were compared, and the group of 35 LEDs with the lowest intensity variations were selected. 

As stated previously, 5 additional LEDs were selected to monitor the degradation of the GdLS and acrylic of the scintillator tanks: four 450~nm and one 390~nm LEDs. Requirements for these LEDs are not as strict as for the 435~nm LEDs because they will be used only for relative measurements: stability of light collection efficiency from these LEDs will be compared to the stability of light collection efficiency from the 435~nm LEDs. 


\subsection{Optical Fibres}
\begin{figure}[t]
     \centering
     \includegraphics[width=0.8\linewidth]{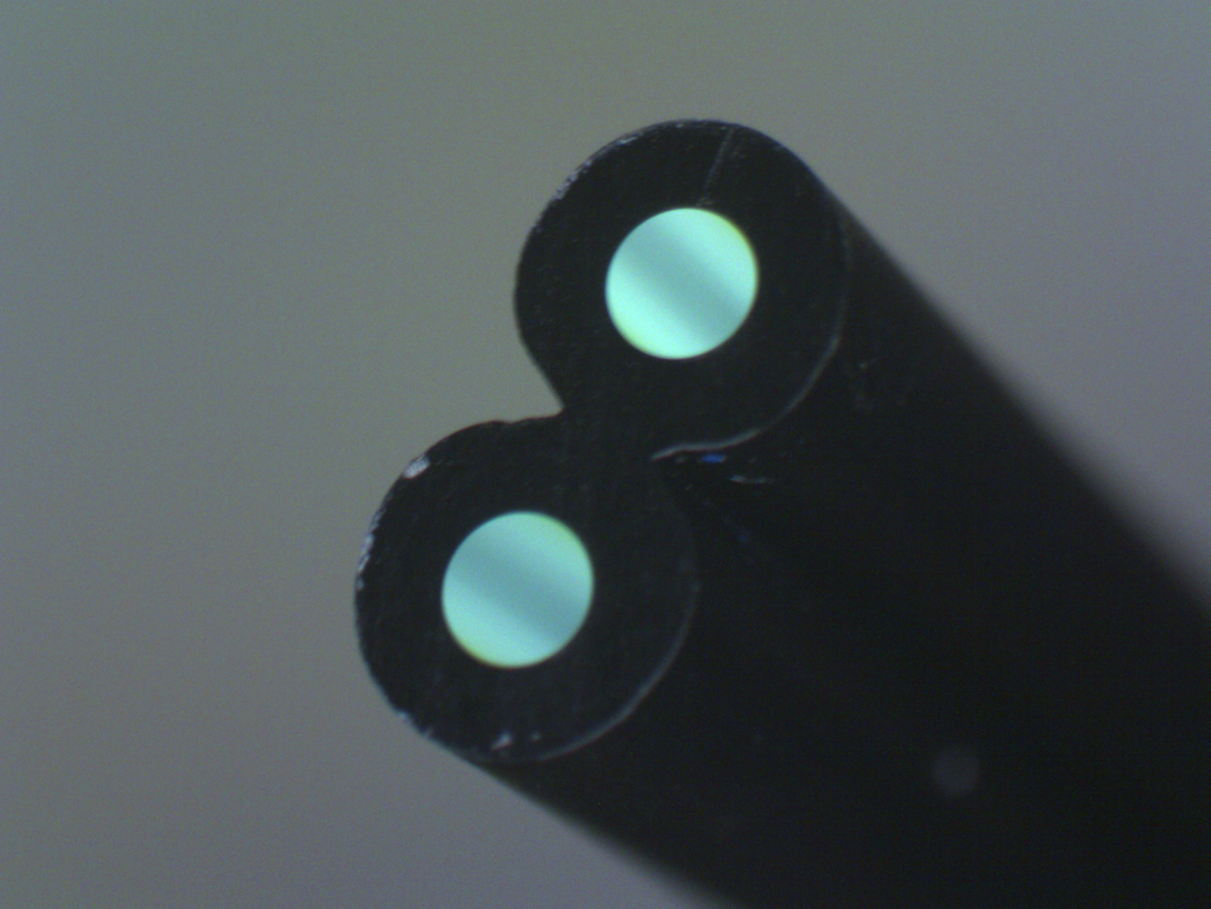}
     \caption{A photo of the fibre end showing the two channels with their polythene jackets after the ice polishing.}
     \label{fig:fibrePhoto}
 \end{figure}

The duplex optical fibres used to direct light to injection points in the water tank are manufactured by Mitsubishi and consist of two polymethyl-methacrylate resin cores, with a refractive index of 1.49, surrounded by a polyethylene jacket \cite{fibreDetails}. The fibre cores measure approximately $\num{980}$~$\mathrm{\mu}$m in diameter and the total diameter of each fibre, including the jacket, is 2.2~mm with the total width of the duplex cable being 4.4~mm as shown in Fig.~\ref{fig:fibrePhoto}. These fibres were selected based on the results of testing by the SNO+ collaboration as our requirements on light transmission and radiopurity are closely aligned \cite{SNO+}.

The fibre injection ends were ice polished at Fermi National Accelerator Laboratory to produce smooth fibre ends, therefore reducing back-scatter and ensuring a uniform beam profile. This process involves submerging the fibre end in liquid nitrogen in order to freeze it in a layer of ice which, not only lubricates the polishing process, but also prevents the cladding and jacket from splitting from the core during polishing~\cite{ice} (see Fig.~\ref{fig:fibrePhoto}).

\begin{figure*}%
  \centering
    \subfigure[][]{%
    \label{fig:beamprofile}%
    \includegraphics[width=0.45\linewidth]{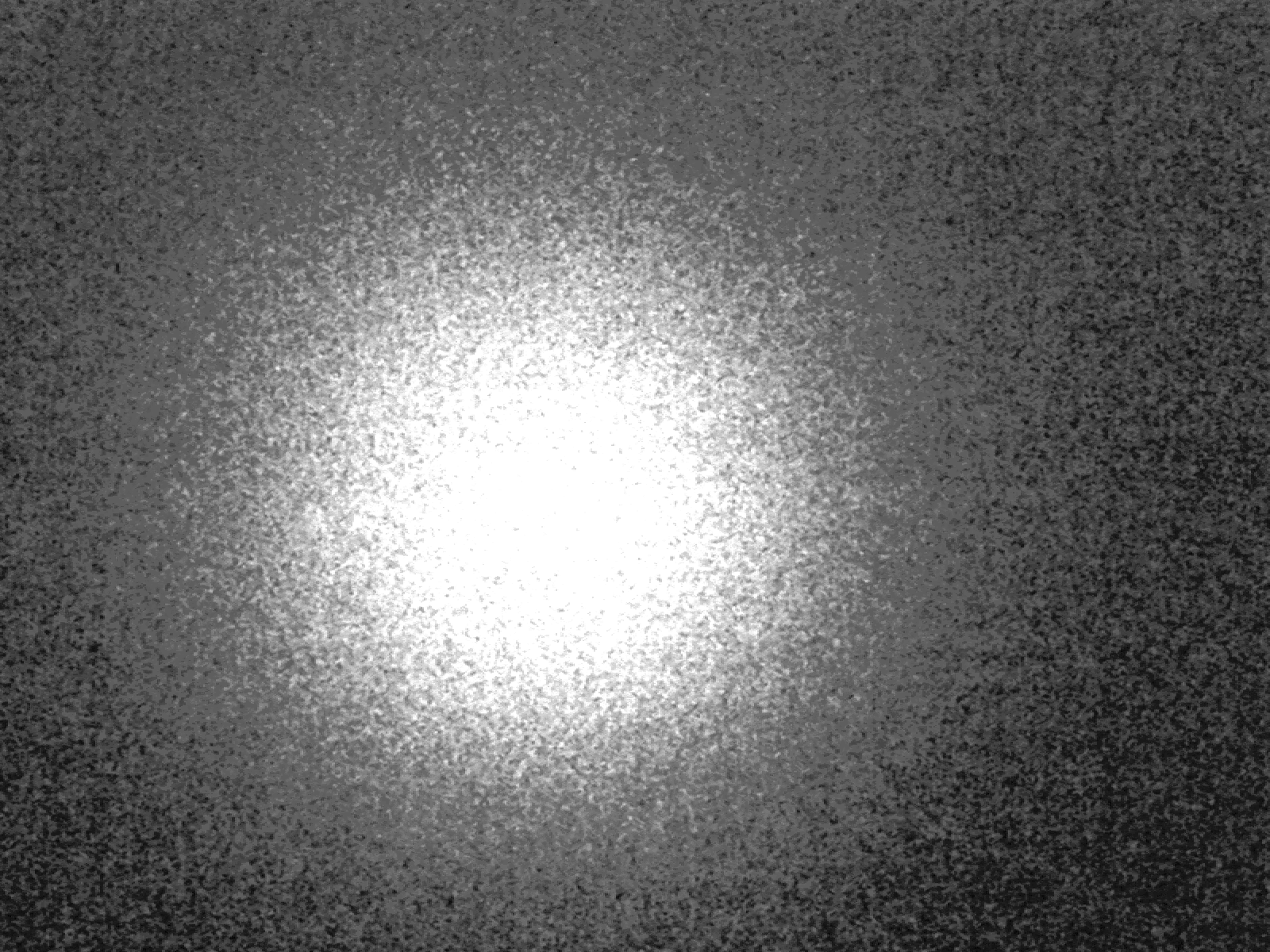}}%
    \hspace{8pt}%
    \subfigure[][]{%
    \label{fig:beamprofileprofile}%
    \includegraphics[width=0.45\linewidth]{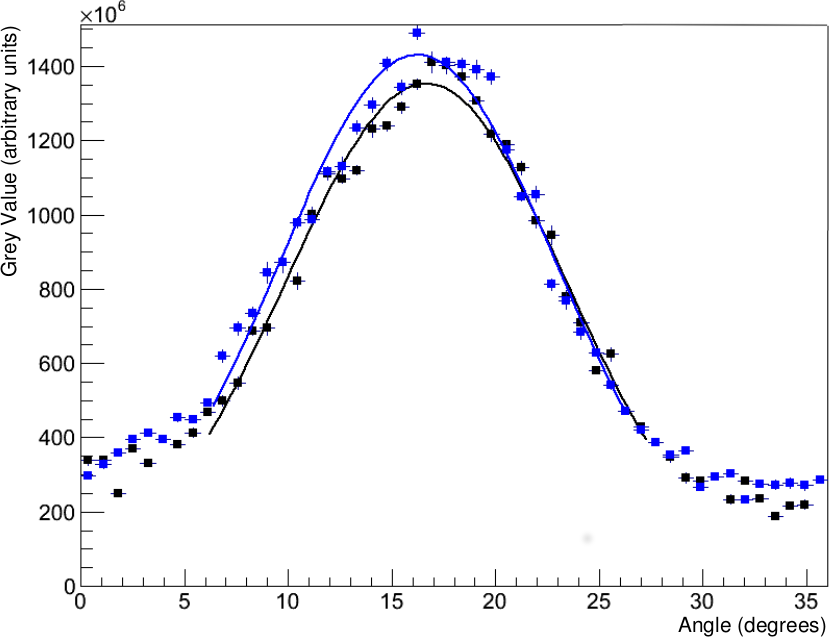}}
      \caption[A set of four subfigures.]{ 
      {Light from a fibre projected on a screen \subref{fig:beamprofile} and
        \subref{fig:beamprofileprofile} the resulting profile plot produced by taking a sample across the middle of the image.The black and blue profiles correspond to the two different channels of the duplex fibre and the slight x-axis offset between them is a result of the 2.2~mm separation between the cores.}}%
\label{fig:LightSpot}%
\end{figure*}

Light profile data was obtained from each of the water tank fibres by shining a light from 435~nm LED down one channel and projecting the emerging light onto a Tyvek\textsuperscript{\textregistered} screen underwater in a dark box. Images such as that in Fig.~\ref{fig:LightSpot} of the light were taken using an endoscope camera and then processed to extract the light profiles for each of the fibres. 

These datas were used as a quality assurance check to ensure that the fibres used have no major defects and the light emitted from them is distributed symmetrically to within an acceptable range. For the fibres to be used for the upward facing injection points, the two channels were compared to ensure they had similar profiles to allow fair comparisons between the two different wavelengths that will be sent through them. In total $\num{49}$ fibres were tested, and the 35 best ones which passed the quality assurance tests were selected for use in the detector. The potential effects of fibre bending on the light profiles were investigated using the same dark box set up by bending a test fibre around discs of varying diameters, from $\num{2}$~cm to $\num{5}$~cm. When comparing the images taken for different diameters, only the smallest disc with a diameter of $\num{2}$~cm showed a small variation in the distribution of the light profile. From these results it was therefore determined that bending the fibres to a diameter of less than $\num{2.5}$~cm should be avoided when they are installed in the detector. Also, the fibre manufacturer measured a power loss of greater than 0.5 \% with a bend radius less than $\num{2.5}$~cm. 

Other quality assurance checks carried out on the fibres included transmission measurements taken with an optical power meter, and the visual inspections of the fibre ends using a microscope. Most of the fibres had near identical power transmission. Any which varied from the average were discarded or kept as a spare. The full lengths of the fibre cladding were then inspected and cleaned before the fibres were packaged in the clean room in sealed bags for shipping. Radon assays were conducted in the manufacture and test premises to ensure acceptable levels of contamination from radon-daughter plateout. 

\subsection{Mechanical Structures}
\begin{figure*}%
  \centering
    \subfigure[][]{%
    \label{fig:Flange}%
    \includegraphics[width=0.45\linewidth]{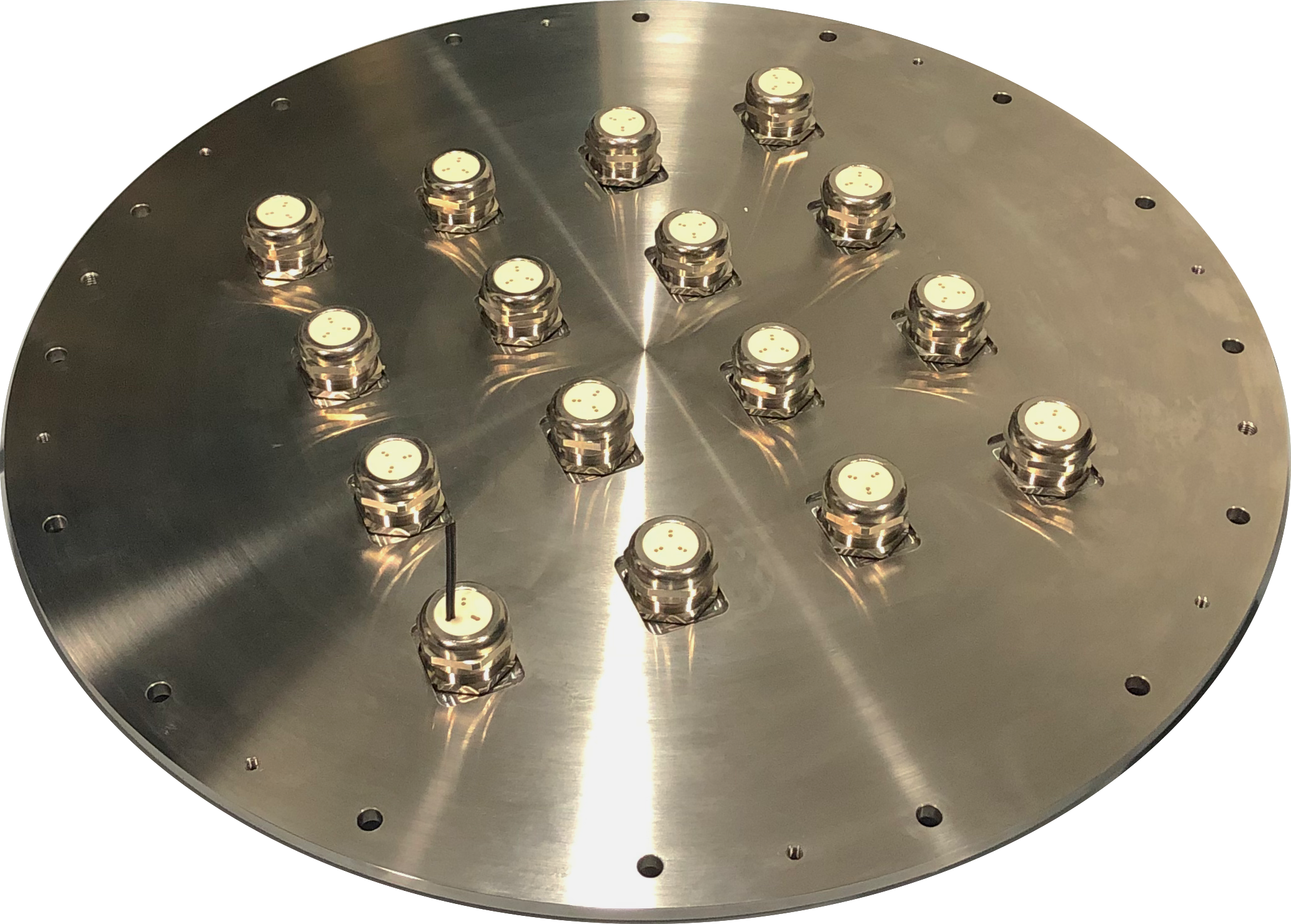}}%
    \hspace{16pt}%
    \subfigure[][]{%
    \label{fig:Bung}%
    \includegraphics[width=0.25\linewidth]{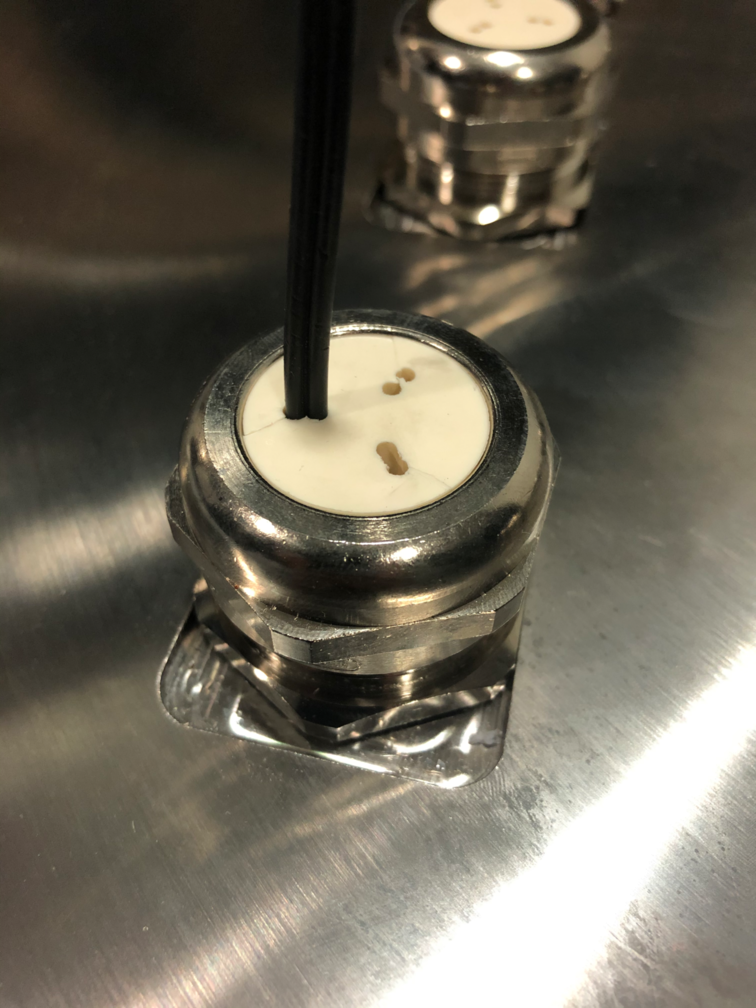}}
      \caption[A set of four subfigures.]{
        \subref{fig:Flange} {Water-tank feedthrough flange.
        \subref{fig:Bung} Feedthrough bung used to allow fibre to pass into the water tank while keeping the water tank light-tight and air-tight.}}%
\label{fig:flangeBung}%
\end{figure*}

A mechanical feedthrough is required to allow the fibres to pass from the electronics rack where the OCS is located into the water tank. Figure~\ref{fig:flangeBung} shows the machined flange with 16 ports which are fitted with rubber bungs to allow the fibres to pass into the water tank while keeping an air-tight and light-tight seal \cite{bung}. There is also a constant nitrogen over-pressure applied above the water level inside the water tank to ensure air does not leak in and introduce radioactive contaminants. 

\begin{figure}
    \centering
    \includegraphics[width=0.8\linewidth]{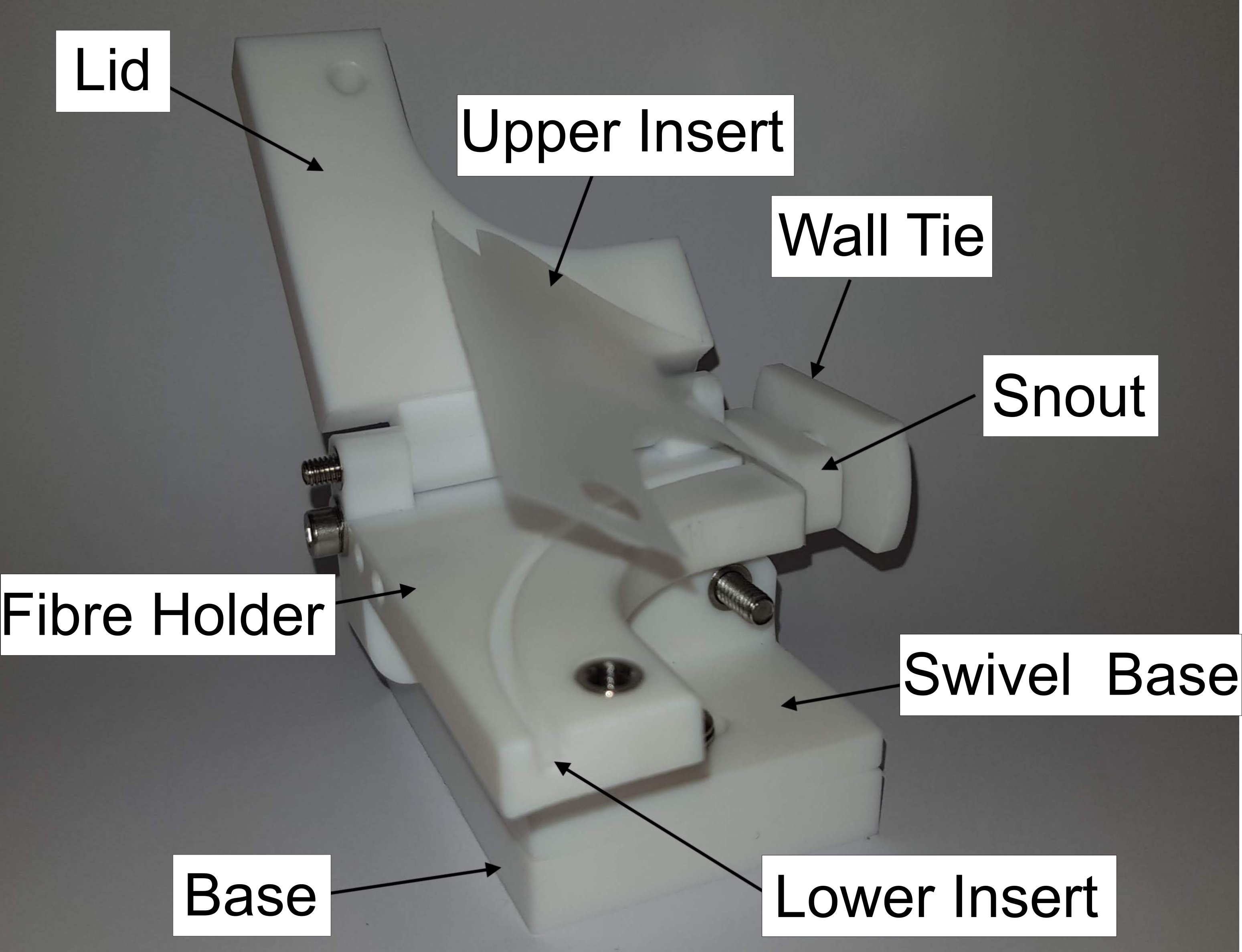}
    \caption{A Fibre Support Structure (FSS) with the lid open. The key components of the FSS are labelled.}
    \label{fig:FSS}
\end{figure}

At each of the 30 injection points on the OD PMT array, the duplex optical fibre will be held in place using a Fibre Support Structure (FSS). The purpose of the FSS is to ensure each fibre is aligned correctly as it points in towards the scintillator tanks and to provide a consistent 50~mm curvature of the duplex optical fibre. Each FSS consists of 26 parts: seven machined PTFE structural components, seven passivated bolts, nine passivated inserts, and two PTFE inserts. A photograph of a FSS with the lid open can be seen in Fig.~\ref{fig:FSS}. The support structures which hold the fibres under the acrylic tanks consist of a custom made stainless steel bracket, which attaches to the acrylic tank platforms, and a PTFE fibre holder which directs the light through holes in the acrylic tank platform. 

Correct alignment of the fibres is of vital importance to allow a realistic comparison between simulations and calibration data. It is for this reason that the FSS are designed to allow setting the fibre direction precisely during installation. The inward facing fibres should point perpendicularly to the water tank wall as well as directing the centre of the beam towards the central axis of the TPC (equivalent to the origin of the water tank coordinate system). To achieve this, two red self-levelling cross laser beams will be positioned equidistantly on either side of the injection point on a custom-made jig with each of the lasers crossing on the face of the liquid scintillator tanks. The FSSs are then aligned to this point by attaching a rear-mounted green laser that points through the fibre injection point and aligning the green laser to the red crosses on the scintillator tank. Once aligned the FSS is tightened to avoid future movement leading to misalignment when the fibre is installed.


\section{Radioactivity Measurements } \label{OCS_RadiationMeasurments}
After the manufacturing of the FSSs was completed, a random selection of six FSSs, as well as the complete set of water tank fibres, was sent to the Boulby Underground Laboratory for radioactivity screening. The results from the screening are shown in Tab.~\ref{table:radio-nuc} along with the estimated activities of the OD components. The activity for the FSSs are calculated by averaging the radioactivity screening results and multiplying by the number of FSSs present in the system.

\begin{table*}[t]
    \centering  
    \small
    \caption{Results from the screening of the OCS components at Boulby Underground Laboratory (shown in bold) in comparison with activities of the radionuclides present in the other OD components as described in the LZ radioactivity and cleanliness control programs article \cite{assay}, where a comprehensive list of the estimated activities for the rest of the detector can be found.}
        
    \begin{tabular}{|x{3cm}|c|c|c|c|c|c|c|x{2.7cm}|} 
    \hline
        \rule{0pt}{3ex}    
     OD Components & Mass (kg) & \multicolumn{6}{c|}{Activity (mBq/kg) } &  Overall Contribution (\%) \\    
 \cline{3-8}
    & & 
    \rule{0pt}{3ex}    
    ${\rm ^{238}U}_e$ & ${\rm ^{238}U}_l$ & ${\rm ^{232}Th}_e$ & ${\rm ^{232}Th}_l$ & ${\rm ^{60}Co}$ & ${\rm ^{40}K}$ & \multicolumn{1}{c|}{ }  \\
    \hline
    OD Tanks & \num{3200} & 0.16 & 0.39 & 0.02 & 0.06 & 0.04 & 5.36 & 3.77\\
    \hline
    Liquid Scintillator & \num{17600} & 0.01 & 0.01 & 0.01 & 0.01 & 0.00 & 0.00 & 0.14\\
    \hline
    OD PMTs & 205 & 570 & 470 & 395 & 388 & 0.00 & 534 & 94.3\\
    \hline
    OD PMT Supports & 770 & 1.20 & 0.27 & 0.33 & 0.49 & 1.60 & 0.40 & 0.59\\
    \hline
    \textbf{Fibre Support Structures} & \textbf{10.5} & \textbf{31} & \textbf{3.0} & \textbf{6.0} & \textbf{3.0} & \textbf{0.5} & \textbf{330} & \textbf{0.77}\\
    \hline
    \textbf{Optical Fibres (inc. couplers)} & \textbf{5.53} & \textbf{91.5} & \textbf{8} & \textbf{10.5} & \textbf{4.5} & \textbf{1.75} & \textbf{302.5} & \textbf{0.45}\\ 
    \hline
    \end{tabular}

    \label{table:radio-nuc}
\end{table*}

These values show that the OCS contributes 1.2\% of the total activity in the OD and therefore satisfies the requirement on radiopurity (see Tab.~\ref{tab:requirements}). The OCS components are also further from the cryostat than the scintillator tanks and therefore have much less of an impact on the number of events in the TPC. The OCS components were thoroughly cleaned before shipping as well as before installation to reduce background contamination further. It is also worth noting that the optical fibres were screened with the metal SMA connector included; these sit outside of the water tank in the experiment so they will not contribute to the detected background events.


\section{Full System Performance}\label{OCS_performance}
\begin{figure}%
  \centering
    \subfigure[][]{%
    \label{fig:ocsDiagram}%
    \includegraphics[width=0.8\linewidth]{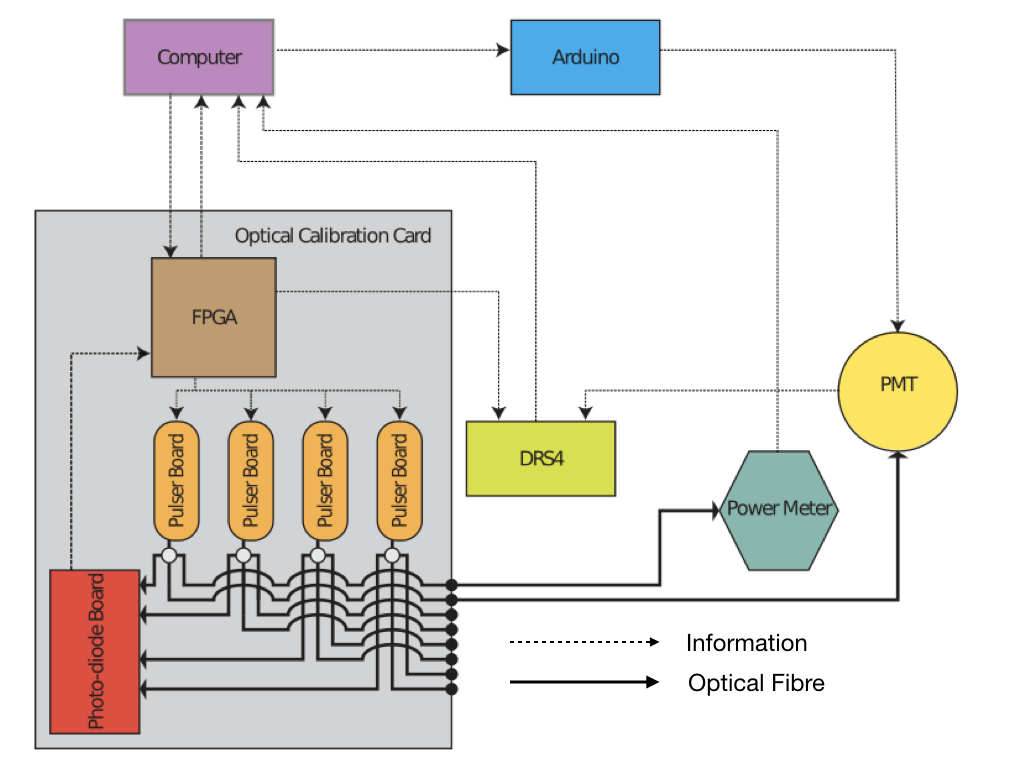}}%
    \hspace{16pt}%
    \subfigure[][]{%
    \label{fig:ocsPicture}%
    \includegraphics[width=0.8\linewidth]{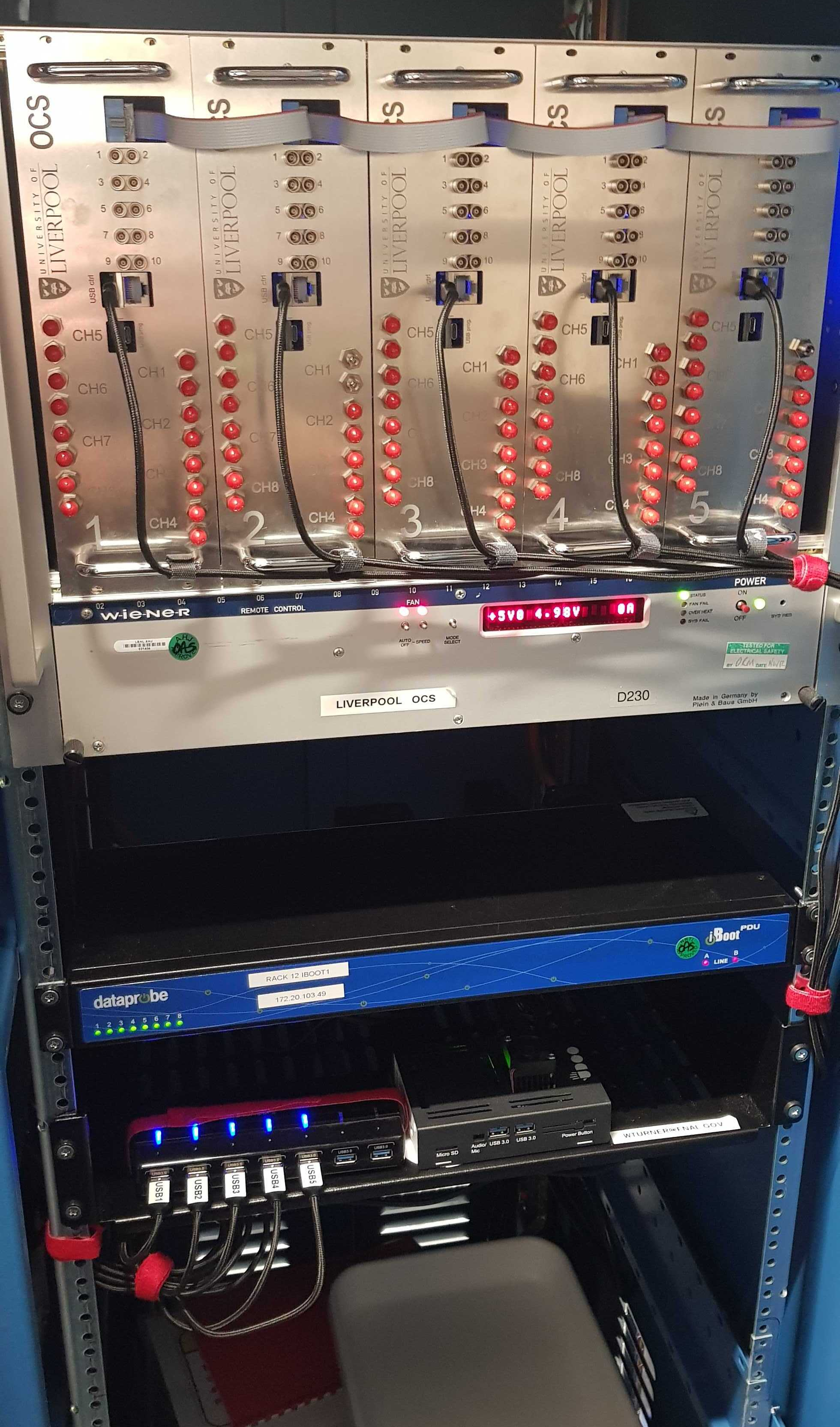}}
      \caption[A set of four subfigures.]{OCS Test Stand:
        \subref{fig:Flange} {An overview of the test stand used at SURF showing the connections made from the OCC to the Thorlab PM100USB Power Meter, PMT, DRS4, Arduino Uno and data taking computer \cite{DRS4} \cite{powerMeter}.
        \subref{fig:ocsPicture} A photograph of the installed Optical Calibration System, showing the VME crate (top) housing the Optical Calibration Cards, each connected by USB to the Optical Calibration Controller (on the shelf).}}%
\label{fig:TestStand}%
\end{figure}

After shipping the OCS to SURF and installing the VME crate in the electronics racks underground, a re-calibration process was carried out to validate the performance of the system. This on-site testing used the same test stand as used previously at Liverpool and provided a final re-calibration for each channel as well as a cross check with the pre-shipment testing. A full system test consisted of going through each channel one by one, and scanning over the range of injectable photons while taking measurements with a power meter (Thorlab PM100USB) and the photomultiplier tube (Hamamatsu H10721)\cite{pmtTestStand}. The power meter output gives the number of photons detected and the photomultiplier tube gives the measured pulse width. Each measurement taken consists of recording the power meter once and recording $\num{4000}$~samples of photomultiplier tube pulses and LED trigger widths which are digitised by a DRS4 Evaluation Board \cite{DRS4}. This measurement is repeated three times at each setting. The $\num{4000}$~samples taken at each measurement are plotted on a histogram and a Gaussian fit is used to determine the mean value and spread of the light intensity for each set point. One of the important parameters determined from the measurements on each channel is the LED trigger pulse width required to produce  a certain number of detected photons. The parameters from this fit can then be used as calibration constants for each channel. 
A schematic of the test stand used for the full system test is shown in Fig.~\ref{fig:ocsDiagram}. A photograph of the installed Optical Calibration System can be seen in Fig.~\ref{fig:ocsPicture}.



\subsection{LED Trigger Width}\label{sectionLEDTW} 
 The trigger sent to the LED has $504$ different indexable widths, which corresponds to a trigger pulse width of $\num{20}$--$\num{60}$~ns. This determines the light output, ranging from $\num{700}$--$\num{700000}$~photons per pulse, while the ability to pulse $\num{1000000}$~photons globally is achieved by pulsing multiple channels simultaneously. The FPGA can change the trigger pulse width with a minimum step of $40$~ps, which is critical to having precise control over the number of photons produced. The trigger width (TW) required for a given number of photons can be calculated using Eq.~\ref{eqn:TriggerWidth}, where $A$, $B$, $C$ and $D$ are channel specific calibration parameters and $N_{ph}$ corresponds to the number of photons which will be injected. This equation was determined empirically.

\begin{equation}
  \begin{split}
    TW &= A + e^{\lambda}, \\
    \lambda &= Bx + Cx^2 + Dx^3, \\
    x &= \text{ln}(N_{ph})
  \end{split}
\label{eqn:TriggerWidth}
\end{equation}

\begin{figure}
    \centering
    \includegraphics[width=0.99\linewidth]{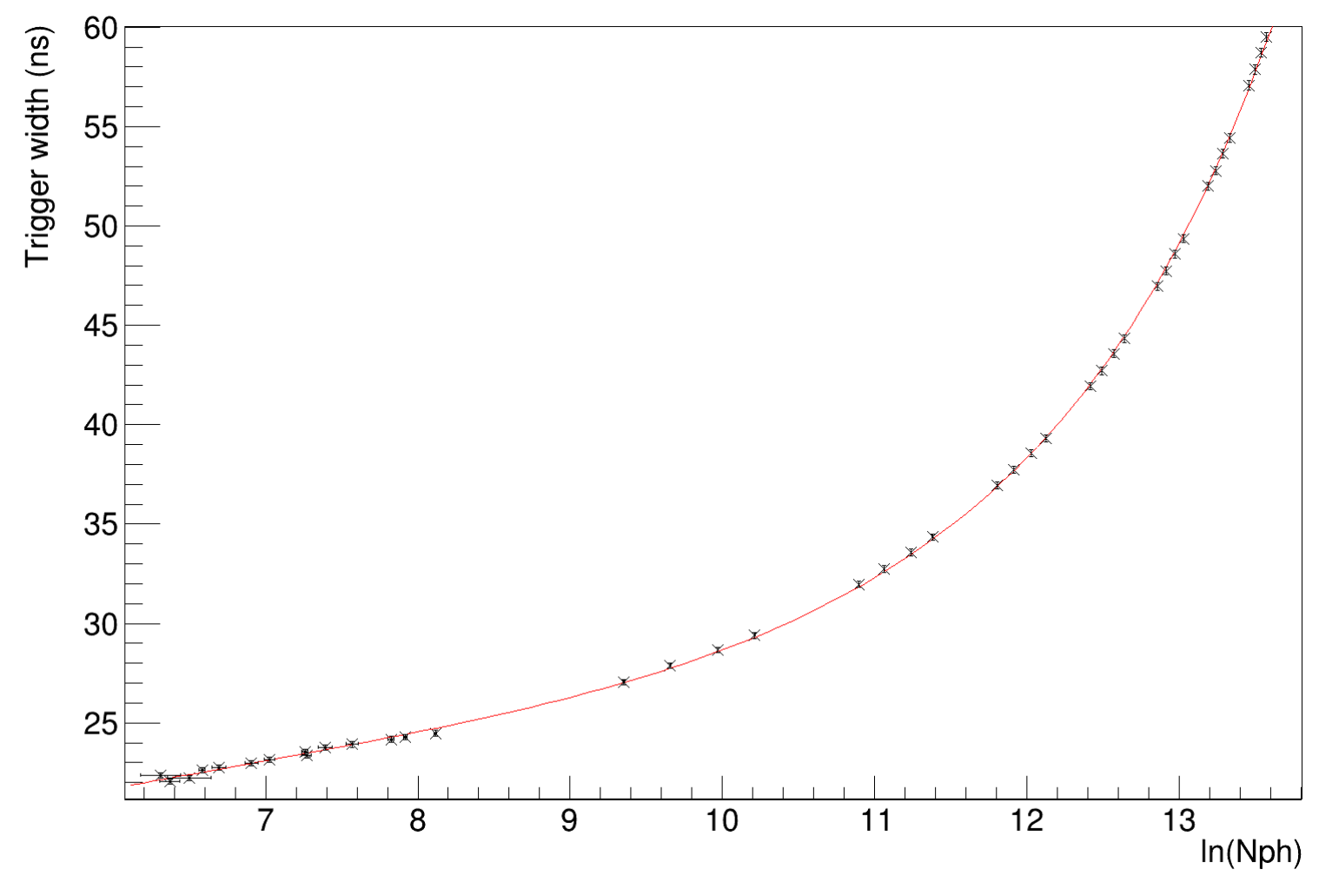}
    \caption{LED trigger width versus the natural logarithm of number of photons per pulse. Equation~\ref{eqn:TriggerWidth} relates these two variables. Each channel is fitted with this function and has a set of parameters for $A$, $B$, $C$ and $D$ which are specific to that channel.}
    \label{fig:TrigWvslnNph}
\end{figure}

An example of the calibration curve is shown in Fig.~\ref{fig:TrigWvslnNph}.

\subsection{PMT Pulse Width}
\begin{figure}
    \centering
    \includegraphics[width=0.99\linewidth]{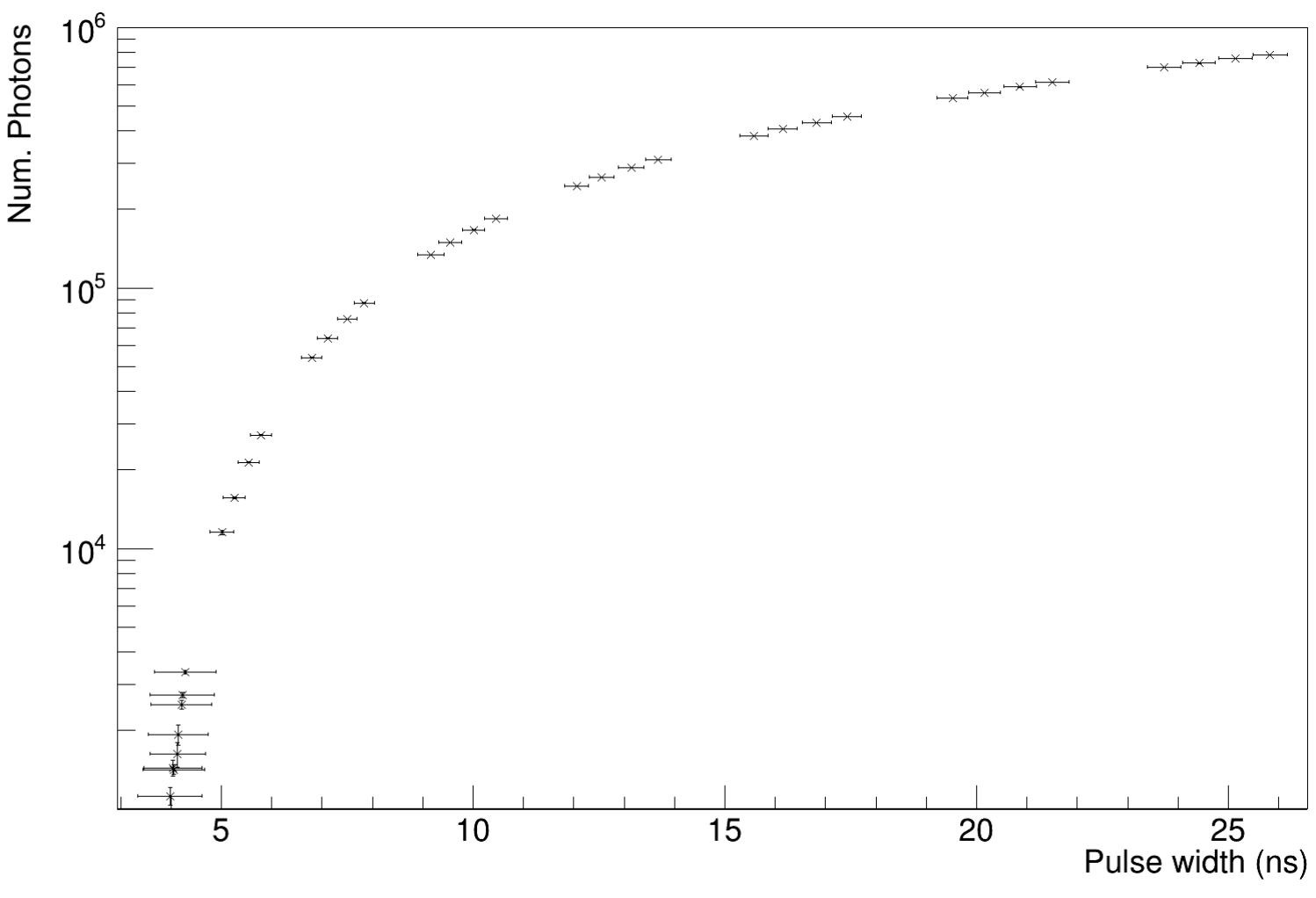}
    \caption{Dependence of the number of photons per pulse on the pulse width measured by the test stand PMT. This plot shows that the system meets the pulse width requirement set by the experiment.}

    \label{fig:NphvsPulseW}
\end{figure}

The channel pulse rate was selected to be $4$ kHz for operation in LZ. However the DRS4 Evaluation Board, which is used to readout the waveforms of the electrical trigger and PMT pulses, can digitise only $300$~Hz with a time resolution of $0.2$~ns. At each setting the collected data is normalised to the number of photons measured by the power meter and analysed to determine pulse width by using the full width at half the maximum peak height. The number of photons detected by the power meter is plotted against the PMT pulse width in Fig.~\ref{fig:NphvsPulseW}. The system satisfies the pulse width requirement where the width of the light pulse must be less than $20$~ns. Fulfilment of this requirement is necessary as the width of the light pulse injected by the OCS must be similar to the width of the light pulse produced in the GdLS or water after being processed by the pulse shaper. 

\subsection{Stability}
\begin{figure}
    \centering
    \includegraphics[width=0.8\linewidth]{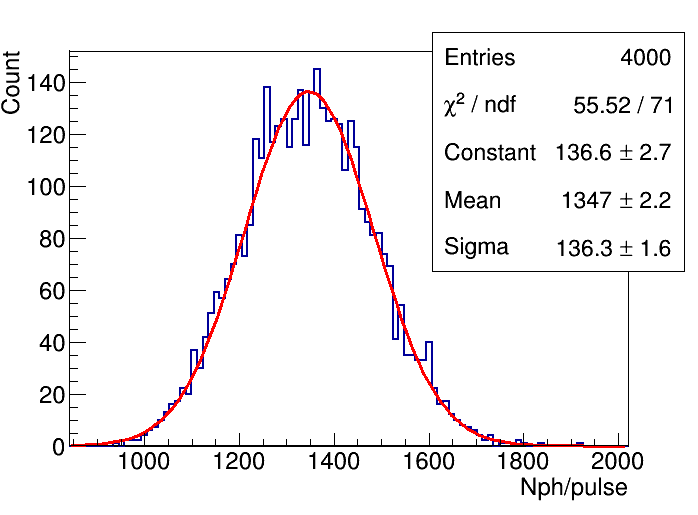}
    \caption{Distribution of numbers of injected photons per pulse at the level corresponding to 150~keV energy deposition in the OD. The resolution 10\%, which is not corrected for the spread due to the PMT measurement, shows that the system meets the pulse width requirement set by the experiment. The resolution is better for higher light intensities.}
    \label{fig:NphSpread}
\end{figure}

Optical calibration and monitoring requires precise knowledge of the numbers of injected photons, especially for measurements related to light collection efficiency. These requirements are reflected in Tab.~\ref{tab:requirements} as pulse-to-pulse variations and variations between calibrations. It is known that electronic components have a temperature dependence which lead to variations in light output from LEDs. Though an effort has been made to use precision components and compensate for these temperature dependencies some residual effects remain. The OCCs have temperature sensors which will allow to study these effects further and implement corrections. Ultimately, light output monitoring has been implemented by integrating over many pulses (photo-diode per channel) and pulse-by-pulse (monitoring PMT for all channels) methods. The pulse-by-pulse resolution at the light intensities corresponding to 150~keV energy depositions in the OD is shown in Fig.~\ref{fig:NphSpread}. The PMT contributes to the width of this distribution; therefore the actual pulse-to-pulse variation is smaller than 10\%, and it improves at higher light intensities.  

The variations of average numbers of injected photons between calibrations will be monitored and corrected for using the photo-diode board and monitoring PMT. 
The YAP:Ce pulser provides continuous gain calibration for the monitoring PMT. Without optical coupling compound, it produced a peak corresponding to $\sim$2000~photons with a resolution of 8\%.  Each OCS calibration will be preceded by a $\sim$2-minute gain calibration of the monitoring PMT allowing to reach 2\% precision (10 minutes are required for the 1\% precision).  

\subsection{Validation}
\begin{figure}%
  \centering
    \subfigure[][]{%
    \label{fig:h700ph}%
    \includegraphics[height=1.5in]{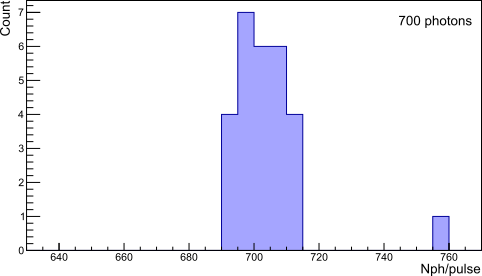}}%
    \hspace{8pt}%
    \subfigure[][]{%
    \label{fig:h1kph}%
    \includegraphics[height=1.5in]{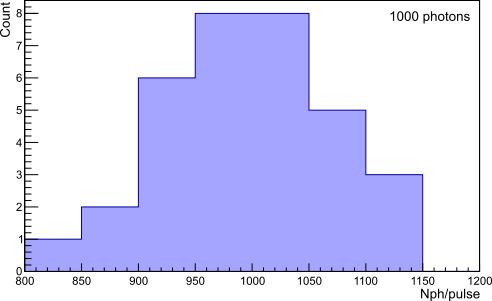}} \\
    \hspace{8pt}%
    \subfigure[][]{%
    \label{fig:h20kph}%
    \includegraphics[height=1.5in]{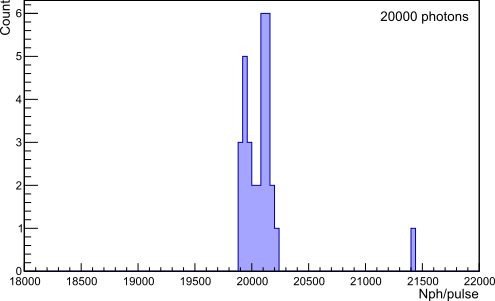}}%
    \hspace{8pt}%
    \subfigure[][]{%
    \label{fig:h50kph}%
    \includegraphics[height=1.5in]{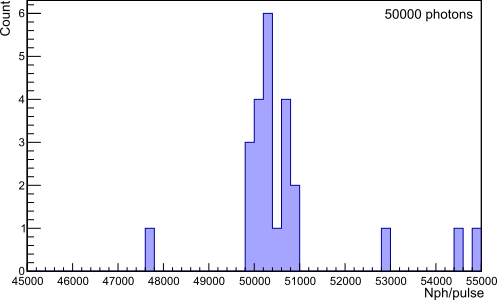}}%
    \hspace{8pt}%

      \caption[]{Distribution of how accurately each channel of the OCS can specifically inject:
        \subref{fig:h700ph} $\num{700}$ photons per pulse;
        \subref{fig:h1kph} $\num{1000}$ photons per pulse;
        \subref{fig:h20kph} $\num{20000}$ photons per pulse; and,
        \subref{fig:h50kph} $\num{50000}$ photons per pulse. This shows the majority of the 40 channels can achieve the desired TW to reach these arbitrary set points (or within 10\%) with the current calibration curves.}%
\label{fig:injectResults}%
\end{figure}

To validate that each channel could hit key calibration points a routine was carried out to see how accurately the channels could hit these photon levels. As described previously, there are certain ranges of TWs reachable for each channel. This test shows if the key calibration points falls within an achievable ranges on each channel. The chosen calibration points were $\num{700}$~photons, $\num{1000}$~photons, $\num{20000}$~photons and $\num{50000}$~photons. Each of the calibration points were entered into Eq.~\ref{eqn:TriggerWidth} along with the parameters for that channel's calibration curve. The resulting trigger width was checked against the reachable range of trigger widths for that channel, taking the closest achievable TW if it was not in an allowable range. The resulting number of photons from this achievable TW was plotted on a histogram. Figure~\ref{fig:injectResults} shows the results of this validation step where all $40$ channels injected light with an accuracy of at least $10\%$ at each of the set points. These results indicate the precision of the OCS internal calibration, which is used to determine the settings necessary to achieve the required number of photons at each set point.

\section{Optical Calibration Procedure}\label{OCS_calibration}
An optical calibration of the OD will be initiated by the LZ Run Control sending the required calibration routine to the Slow Control system. The Slow Control is a SCADA platform which utilises Ignition \cite{LZTDR} and communicates with LZ's subsystems via the MODBUS protocol \cite{LZTDR}. The Slow Control system will store and manage the calibration constants needed for the OCS in a MySQL database.
The OCS control commands will pass from the Slow Control to the OCS controller which manages the communication with the master OCC. The OCS controller will also measure OCC temperatures and monitor photo-diode response.

\begin{figure}[ht!]
	\centering
	\includegraphics[width=0.8\linewidth]{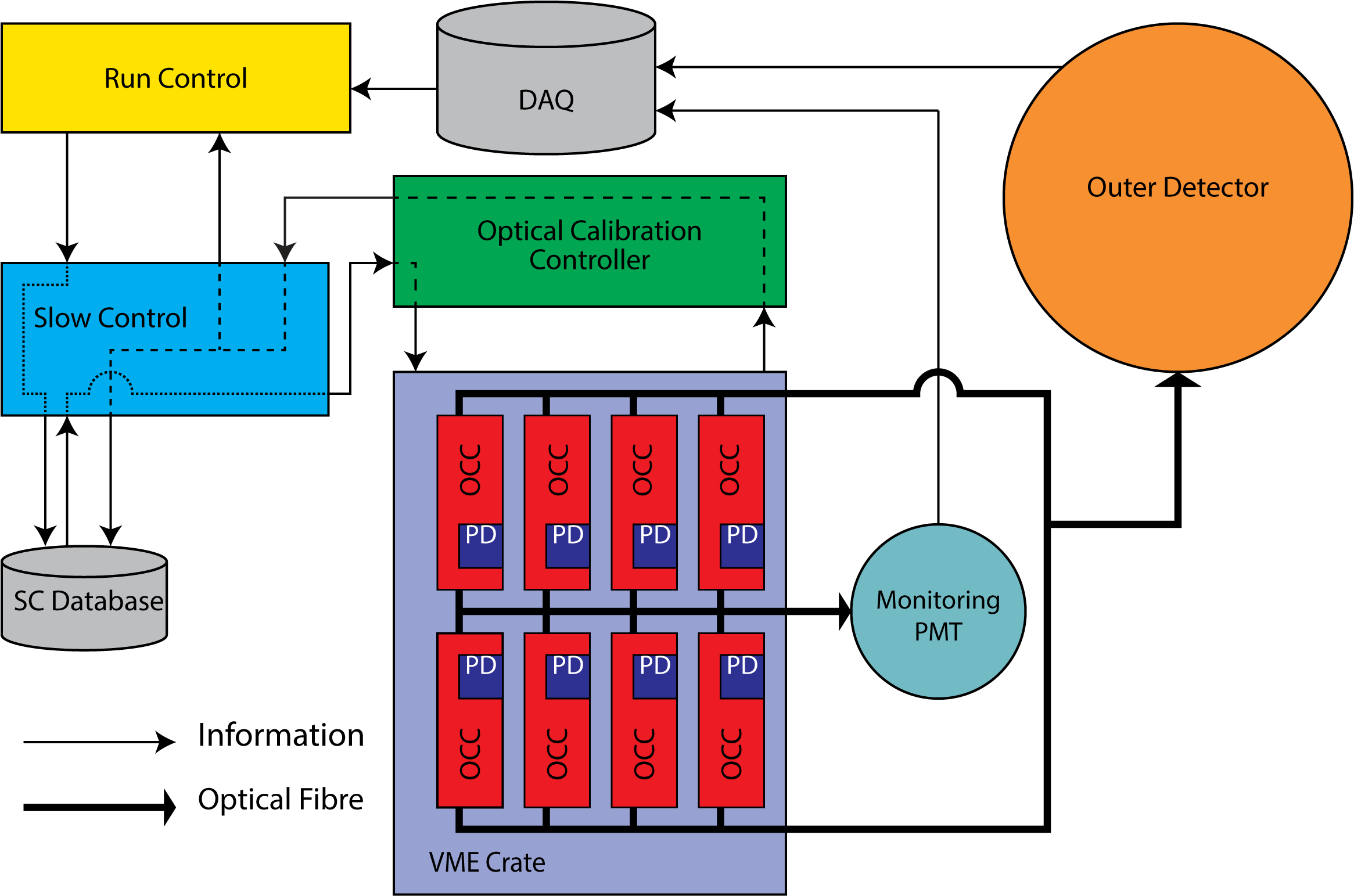}
	\caption{A schematic showing the order in which an OD optical calibration will be completed.}
	\label{fig:OCS_Flow}
\end{figure}

The monitoring PMT will be connected to the LZ data acquisition system. An analysis of OD-PMT data against the monitoring PMT data will be used to ensure that the OD-PMTs are correctly calibrated and identify any drift in performance. A schematic of the OCS communication with the LZ Run and Slow Control can be seen in Fig.~\ref{fig:OCS_Flow}.

\section{Conclusions}
The ability to perform an accurate calibration of the OD is critical to ensure the effectiveness of the veto system used by LZ. The OD is essential for understanding and vetoing the signal-like neutron background as well as numerous gammas originating inside and outside of LZ. The OCS discussed in this paper has the capability to precisely inject a known number of photons ranging from $\num{700}$~photons up to $\num{50000}$~photons covering the energy range of the background signals. Light will be injected into the OD using an LED-driven system with $30$~duplex optical fibres mounted within the array of PMTs and five duplex optical fibres mounted beneath the tanks pointing upwards into acrylic tanks. The injection points situated within the array will maintain the calibration of PMTs and monitor their performance. The five bottom injection points will monitor the light transmission of the acrylic tanks and the degradation of the liquid scintillator over time. A correctly calibrated OD ensures that the experiment will be able to meet its projected sensitivity.



\section{Acknowledgements}
We would like to thank our collaborators in the LZ dark
matter search experiment for their continuous support, in particular, H.~Kraus from Oxford, H.~M.~Ara\'ujo from Imperial College London,  M.G.D. van der Grinten and R. Preece from Rutherford Appleton Laboratory, H.~Nelson from U.C. Santa Barbara for useful discussions on the OCS design and development; our colleagues at Brandeis University for tests of the OCS parts; A.~Kaboth from Royal Holloway University of London, R.~Mannino from University of Wisconsin Madison, 
A.~Kamaha from University at Albany, B.~Penning from University of Michigan, B.~L\'{o}pez Paredes from Imperial College London for comments on this manuscript.
We wish to thank H.~Lippincott (now at U.C. Santa Barbara) and E.~Hahn from FNAL for the fibre ice polishing.
Thanks are due to M.~Whitley, P.~Cook and the staff working in the University of Liverpool Detector Fabrication Facility and Advanced Materials Laboratory for hardware manufacture. 
We acknowledge contributions from University of Liverpool students T.~Carter, L.~Hawkins, A.~Hibbert, B.~Philippou.
We thank N.~McCauley, J.~Rose, A.~Pritchard and L.~Anthony from the University of Liverpool Hyper Kamiokande group for their help and design discussions. We acknowledge additional support from the Boulby Underground Laboratory for organising and carrying out the radiation screening required for the OCS components. 
We acknowledge the LZ collaboration for providing the BACCARAT simulation package which allowed the simulations mentioned in this article to be carried out. With thanks going to S.~Shaw from U.C. Santa Barbara for help writing the simulations for the OCS.  
The assistance of Sanford Underground Research Facility and its personnel in providing physical access and general logistical and technical support is acknowledged.
This work was supported by the U.K. Science \&
Technology Facilities Council (STFC) under award numbers ST/S000879/1 and ST/M003639/1, and by PhD studentships 
ST/N504142/1 (AB), ST/R504920/1 (AB) and ST/S505547/1 (EF).


\section*{References}

\bibliography{citations}
\end{document}